\documentclass[%
 aip, apl,
 amsmath,amssymb,
 reprint,%
]{revtex4-1}

\usepackage{graphicx}% Include figure files
\usepackage{dcolumn}% Align table columns on decimal point
\usepackage{bm}% bold math
\usepackage{subcaption} % for using subfigure.
\usepackage{hyperref} % for making contents and \ref be hyperlinks

\begin{document}

\title{\bf Simulation Study of the Influence of Experimental Variations on the Structure and Quality of Plasma Liners}

\author{
\bf Wen Shih$^1$, Roman Samulyak$^{1,2}$, Scott C. Hsu$^3$, Samuel J. Langendorf$^3$, Kevin C. Yates$^4$,  and Y.C. Francis Thio$^5$
\\ \it $^{1}$Department of Applied Mathematics and Statistics, 
       Stony Brook University, Stony Brook, NY 11794, USA
\\ \it $^{2}$Computational Science Initiative, Brookhaven National Laboratory, Upton, NY 11973, USA 
\\ \it $^3$Physics Division, Los Alamos National Laboratory, Los Alamos, NM 87545, USA
\\ \it $^4$Department of Electrical and Computer Engineering, University of New Mexico, Albuquerque, NM 87131, USA
\\ \it $^5$HyperJet Fusion Corporation, Chantilly, VA 20151, USA}

%%%%%% abstract begins %%%%%% [
\begin{abstract}
Simulation studies of a section of a spherically imploding plasma liner, formed by the merger of six hypersonic plasma jets, have been performed at conditions relevant to the Plasma Liner Experiment (PLX) [S. C. Hsu et al., IEEE Trans. Plasma Sci.~{\bf 46}, 1951 (2018)].
The main aim of simulations was the sensitivity study of the detailed structure of plasma liners and their global properties to experimental mass variations and timing jitter across the six plasma jets. 
Experimentally observable synthetic quantities have been computed using simulation data and compared with the available experimental data. Simulations predicted that the primary oblique shock wave structure is preserved at small experimental variations.  At later phases of the liner implosion, primary shocks and, especially, secondary shocks are more sensitive to experimental variations. 
These conclusions follow from the simulation data as well as comparisons between synthetic and experimental interferometry and visible images.
Small displacements of the shock wave structures may cause significant changes in the synthetic interferometer data at early time.
Our studies also showed that the global properties of the plasma liners (averaged Mach number and averaged ram pressure along leading edges of plasma liners) are not very sensitive to experimental variations. 
Simulation data of the liner structure were largely confirmed by the PLX experimental data.
\end{abstract}
%%%%%% abstract ends %%%%%% ]

\maketitle

\section{Introduction}

The main goal of this work is to perform simulation studies of the effect of experimental imperfections on the structure and quality of plasma liners for 
plasma-jet-driven magneto-inertial fusion (PJMIF) and to validate simulations using experimental data.
In the PIMIF concept,  a spherically symmetric plasma liner, formed by the merger of a spherically distributed array of highly supersonic plasma jets, implodes on a magnetized target and compresses it to the fusion conditions \cite{Thio99,Hsu12,Knupp14}.  
Solving the so-called "stand-off problem" of solid liners, i.e. difficulties in assembling solid liners from a sufficient distance from the fusion hot spot, is among the main potential advantages of PJMIF compared to the solid liner-driven MIF.
Simulation studies reported in this work have been performed in support of  the Plasma Liner Experiment -- ALPHA (PLX-$\alpha$) project, the primary objective of which is to form and study a spherically imploding plasma liner with at least 36 and up to
60 merging plasma jets that are launched by an array of coaxial plasma guns from the periphery of a 2.74-m-diameter vacuum chamber. The first series of experiments operating with 6 jets have been recently completed  \cite{PLX18}.

Plasma liners and the corresponding compression of plasma or gas targets have been studied by a number of analytic models and 1D hydrodynamic codes 
\cite{Thio02,Parks08,Hsu12,Knupp14,SamParWu10,KimSam12,Awe11,HsuLang18}. These works focused on the study of averaged properties of idealized plasma liners, their ram pressures, Mach numbers, and hydrodynamic efficiencies, the dependence of these quantities on material parameters, atomic processes and radiation, and the compression and fusion gains of plasma targets. As 3-dimensional processes that contribute to deterioration of quality of plasma liners are neglected in the spherically symmetric approximation, such works evaluate only an upper limit of the liner-target performance for each study case.
First 3D simulations of detailed structure and properties of  argon plasma liners obtained by the merger of 60 jets, reported in \cite{KimZhaSam13}, predicted a cascade of oblique shock waves during the jet merger process and  their role in seeding non-uniformities in plasma liners.
This work continues detailed numerical simulations of plasma liners at conditions closely related to PLX, confirming previous predictions, presents comparisons with experiments, and  studies the sensitivity of the structure and properties of liners to variations in experimental conditions. It must be emphasized that the present experiments are not intended to demonstrate fusion-worthy liner quality and uniformity, which will most likely require smaller merging angles between plasma jets and many more jets (e.g., hundreds of jets over a sphere of 3 to 4-m radius).  The purpose of this work is to benchmark our models and codes such that they may be used to investigate and help design follow-on experiments with more jets at smaller merging angles.

The rest of the paper is organized a follows. Section II describes numerical models and simulation setups. Analysis of numerical simulations that model two experimental scenarios with different values of variations of the initial mass of gas in plasma guns and jet velocities, and their role on local and global properties of plasma liners are presented in Section III. Comparison of simulations with experiments is the subject of Section IV. Finally, we conclude the paper with a brief summary of our results and perspectives for the future work.

\section{Numerical Models and Simulation Setup}

Simulation results reported in this paper have been performed using FronTier \cite{FronTier},  a code for multiphysics
simulations of compressible hydrodynamic and low magnetic Reynolds number MHD flows that uses the method of front tracking for accurate resolution of material interfaces in multiphase flows. 
Plasma liner-specific physics models in FronTier include an equation of state for high-Z materials  with atomic physics transformations, based on the Zeldovich average ionization model \cite{KimSam12}, and
a radiation model in the thin optical limit approximation, in which the Plank's emission opacity is calculated using data generated by the  PROPACEOS code (see Appendix A of \cite{PROPACEOS}).
Other details of numerical models and the use of FronTier for the simulation of PJMIF processes, including 3D simulations of the formation and implosion of plasma liners are described in \cite{KimZhaSam13,Shih18}.

\begin{figure}[h!] 
      \includegraphics[width=0.5\textwidth] % [width=0.7\textwidth]{ram_pres_radial.jpg} 
        {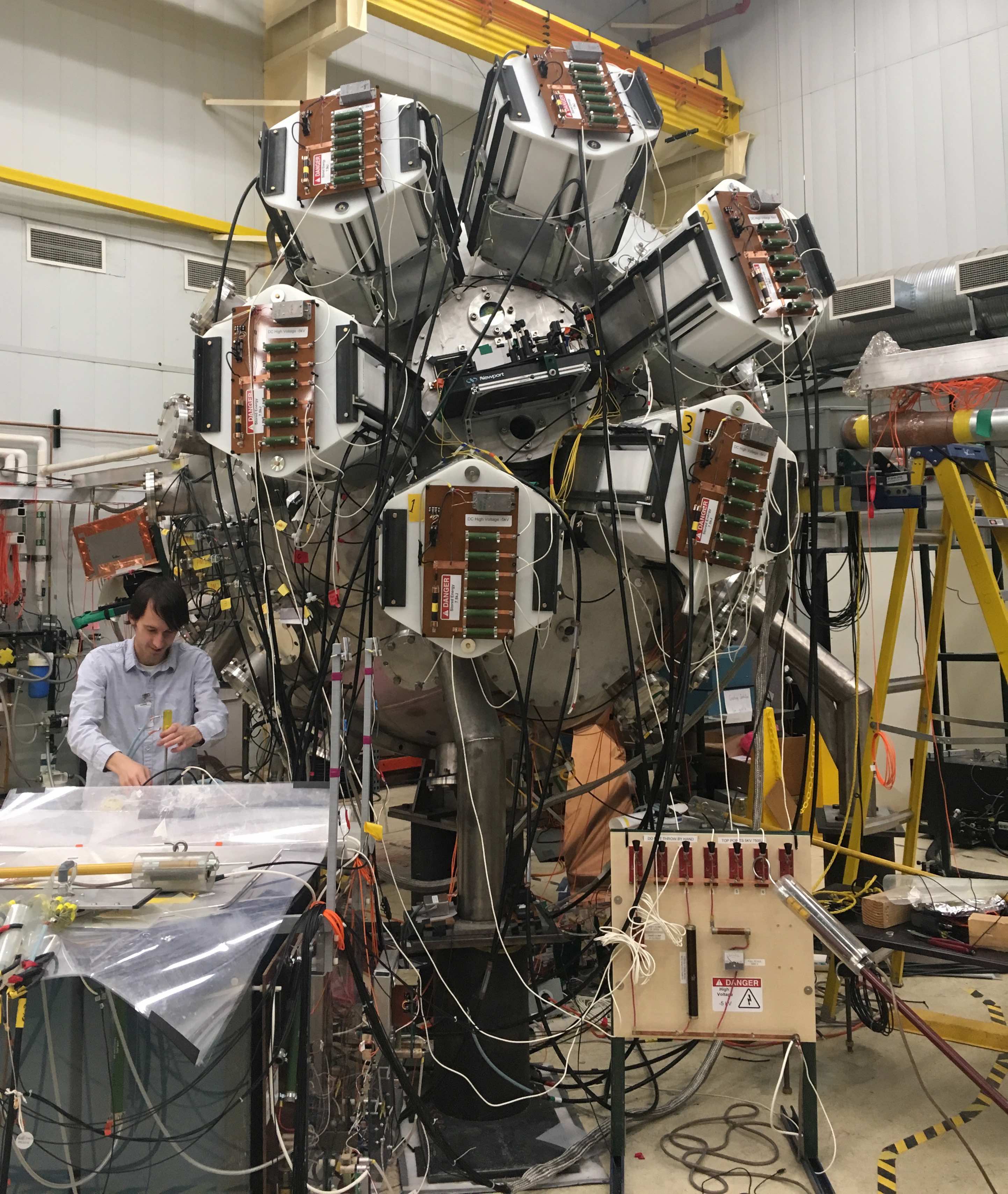} % size and path
    \caption{
      \raggedright
      Photograph of the PLX experimental chamber at LANL showing the location of six plasma guns. The mingle section in the center of the jet array contains multi-chord interferometer hardware.
    } % Description of the figure and the figure will be auto-numbered
    \label{fig:PLX_exp} 
 \end{figure}

In this work, we study the merging process of 6 plasma jets and  the formation and implosion of a section of the plasma liner  by closely following the setup of the PLX-$\alpha$ experiment at Los Alamos. 
While the experimental chamber is capable of holding 60 plasma guns in a 4$\pi$ configuration, the first series of experiments operated with 6 closest plasma guns, forming a small section of the spherical liner. The location of plasma guns in the experimental device is shown in
Figure \ref{fig:PLX_exp}.
The simulation parameters are as follows. The radius of the chamber is 130 cm. The initial argon plasma jets have density $\rho = 2 \times 10^{16}$ 1/cm$^3$ 
$ = 1.327 \times 10^{-6}$ g/cm$^3$, temperature $T=2.5$ eV, and velocity 34.7 km/s. 
The ambient vacuum is modeled as rarefied gas with density $\rho_0 \sim 10^{-9}$ 
g/cm$^3$ and pressure $P_0 \sim 10^{-6}$ bar. The diameter and length of the jets are 
8.5 cm and 10 cm, respectively.

Ideally, identical plasma jets should be shot synchronously from the plasma guns at the periphery of the PLX chamber. 
In practice, however, there are variations in the mass injected and precise trigger times of the six guns.
We are interested in the understanding of how such variations change the internal structure of plasma liners as well as their global properties. We study two different experimental scenarios. In the first scenario, initial masses of argon jets and their shot times experience random
variations.  In particular, we study random variations of jet masses within $\pm 10\%$ and $\pm 2\%$, together with random variations of shot time within $2\mu s$, $100 ns$, and $50 ns$. Table \ref{tab:perturbed_set} summarizes parameters for this series of simulations.  For a better understanding of the details of numerical results, we list exact parameters for each jet which were generated by random numbers.
    
 In the second experimental scenario, the plasma guns are shot perfectly synchronously but the amount of argon gas in the plasma guns varies by $\pm 5\%$. We assume that plasma guns convert the same amount of energy into the kinetic energy of plasma jets, and compute the corresponding variations in the initial jet velocity. Specific initial parameters are presented in Table \ref{tab:perturbed_second_set}. Both simulation series are compared with the idealized simulation that implements identical initial states for each plasma gun.

    %%%%%%%%%%%%%%%%%%%%%%%%%%%%%%%%%%%%%
    % begin table: flow %%%%%%%%%%%%%%%%% [
    \begin{table}[t] % [H]: fix talbe position, [h]: try best to fix position
    \begin{center}
      %%%%%%%%%%%%%%%%%%%%%%%%%%%%%%%%%%%%%
      \begin{subtable}{0.49\textwidth}
      \begin{center}
        \begin{tabular}{c | c | c | c | c | c | c} % 7 cloumns 
        \hline 
        \hline 
                               & Jet 1 & Jet 2 & Jet 3 & Jet 4 & Jet 5 & Jet 6 \\
        \hline 
        Change in mass [\%]     & -8.61 & 6.72  & 6.31  & -3.39 & -6.69 & -5.09 \\
        \hline 
        Change in time [$\mu$s] & -1.9  & -0.9  & -0.3  & 0     & -0.2  & -1.87\\ 
        \hline 
        \hline 
        \end{tabular}
      \subcaption{
        \raggedright
        Variation 1: Random variations in mass of argon within 10\% with 
        random errors in gun shot times within 2 $\mu$s.
      } % descirption of the table and table will be auto-numbered
      \label{tab:perturbed_set_a} % must be after caption
      \end{center}
      \end{subtable}
      %%%%%%%%%%%%%%%%%%%%%%%%%%%%%%%%%%%%%
      \begin{subtable}{0.49\textwidth}
      \begin{center}
        \begin{tabular}{c | c | c | c | c | c | c} % 7 cloumns 
        \hline 
        \hline 
                           & Jet 1 & Jet 2 & Jet 3 & Jet 4 & Jet 5 & Jet 6 \\
        \hline 
        Change in mass [\%] & 9.65  & -9.27 & 4.93  & 1.37  & 2.58  & -9.98 \\
        \hline 
        Change in time [ns] & -96   & 100   & 39    & 62    & -97   & -9\\ 
        \hline 
        \hline 
        \end{tabular}
      \subcaption{
        \raggedright
        Variation 2: Random variations in mass of argon within 10\% with 
        random errors in gun shot times within 100 ns.
      } % descirption of the table and table will be auto-numbered
      \label{tab:perturbed_set_b} % must be after caption
      \end{center}
      \end{subtable}
      %%%%%%%%%%%%%%%%%%%%%%%%%%%%%%%%%%%%%
      \begin{subtable}{0.49\textwidth}
      \begin{center}
        \begin{tabular}{c | c | c | c | c | c | c} % 7 cloumns 
        \hline 
        \hline 
                           & Jet 1 & Jet 2 & Jet 3 & Jet 4 & Jet 5 & Jet 6 \\
        \hline 
        Change in mass [\%] & 1.44  & -0.10 & 1.10  & 0.29  & -0.80 & -0.43 \\
        \hline 
        Change in time [ns] & -58   & 43    & -37   & 99    & -55   & -37\\ 
        \hline 
        \hline 
        \end{tabular}
      \subcaption{
        \raggedright
        Variation 3: Random variations in mass of argon within 2\% with 
        random errors in gun shot times within 100 ns.
      } % descirption of the table and table will be auto-numbered
      \label{tab:perturbed_set_c} % must be after caption
      \end{center}
      \end{subtable}
      %%%%%%%%%%%%%%%%%%%%%%%%%%%%%%%%%%%%%
      \begin{subtable}{0.48\textwidth}
      \begin{center}
        \begin{tabular}{c | c | c | c | c | c | c} % 7 cloumns 
        \hline 
        \hline 
                           & Jet 1 & Jet 2 & Jet 3 & Jet 4 & Jet 5 & Jet 6 \\
        \hline 
        Change in mass [\%] & 0.09  & -0.76 & 0.62  & -0.21 & 1.35  & -1.20 \\
        \hline 
        Change in time [ns] & -5    & 50    & 43    & 37    & -27   & 18\\ 
        \hline 
        \hline 
        \end{tabular}
      \subcaption{
        \raggedright
        Variation 4: Random variations in mass of argon within 2\% with 
        random errors in gun shot times within 50 ns.
      } % descirption of the table and table will be auto-numbered
      \label{tab:perturbed_set_d} % must be after caption
      \end{center}
      \end{subtable}
      %%%%%%%%%%%%%%%%%%%%%%%%%%%%%%%%%%%%%
    \caption{
      \raggedright
      Changes of initial jet parameters in experimental scenario 1. 
    } % descirption of the table and table will be auto-numbered
    \label{tab:perturbed_set} % must be after caption
    \end{center}
    \end{table}

\begin{table}
\begin{center}
\begin{tabular}{c | c | c | c | c | c | c} % 7 cloumns 
\hline 
\hline 
                               & Jet 1 & Jet 2 & Jet 3 & Jet 4 & Jet 5 & Jet 6 \\
\hline 
        Change in mass [\%]     & -2.15 & -1.28  & 3.97  & -2.48 & 0.22 & 1.44 \\
%\hline 
%        Density [g/cc]  & 1.2985E-06 &1.3100E-06 & 1.3797E-06 & 1.2941E-06 & 1.3299E-06 & 1.3461E-06 \\
\hline 
        Velocity [km/s] & 35.0  & 34.92  & 34.03  & 35.14  & 34.66  & 34.45 \\
\hline 
\hline 
\end{tabular}
\caption{
      \raggedright
        Changes of initial jet parameters in experimental scenario 2. } % descirption of the table and table will be auto-numbered
\label{tab:perturbed_second_set} % must be after caption
\end{center}
\end{table}

\section{Numerical Results}

\subsection{Formation of plasma liner by merging jets}

Consider an array of 6 supersonic plasma jets, shot by synchronized plasma guns from the periphery of the PLX chamber. We start with a qualitative discussion of important processes occurring during the formation and implosion of liners that strongly influence their quality. As jets propagate towards the chamber center, they expand and cool down due to the adiabatic expansion and radiation losses.  As the initial jet length is close to its diameter, the jets look spherical in shape in the middle of the chamber. PLX operates with somewhat longer plasma jets compared to that used in simulations. But since the main contribution of the liner to the target compression, to be explored in the future, is provided by the front part of the plasma jets, the reduction of the jet length in numerical simulations to 10 cm does not lead to essential differences in the estimation of main liner properties while significantly reduces the computational time. Justification for the  above statement can be found in \cite{SamParWu10}, which shows that using liners of increased thickness does not lead to higher target compression rates but reduces instead the hydrodynamic efficiency of liners, defined as the ratio of the internal energy of the compressed target to the initial energy of the liner. As jets collide with each other, strong oblique shocks are formed that affect the quality of plasma liners.

\begin{figure}[h] % flow mode; [H]: fix the position of tables, graphs
    %\begin{center}
      \includegraphics[width=0.4\textwidth] % [width=0.7\textwidth]{ram_pres_radial.jpg} 
        {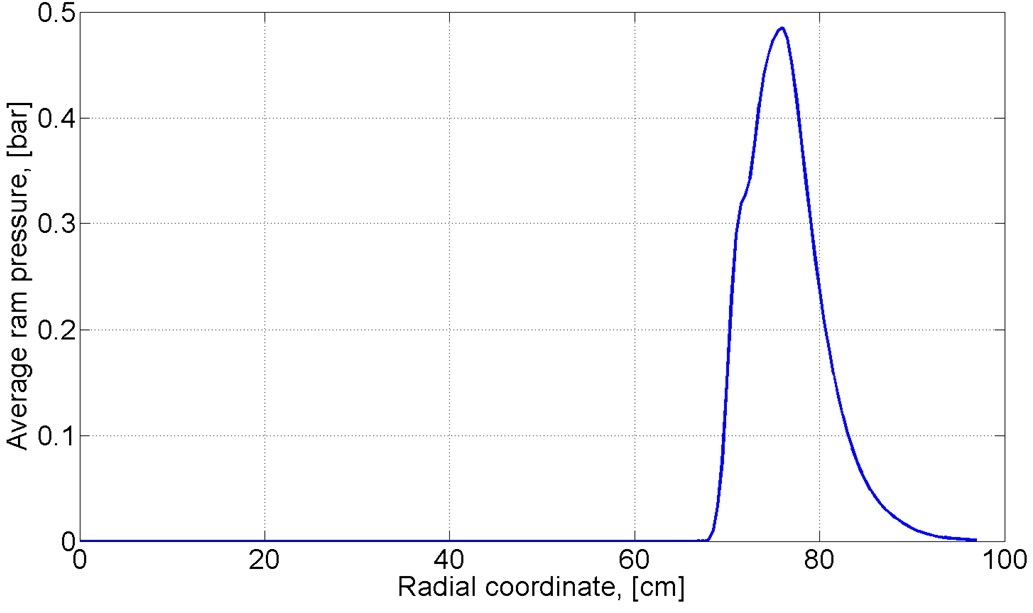} % size and path
    \caption{
      \raggedright
      Radial distribution of the averaged ram pressure in the liner formed by the 
      merger of 6 jets at time 17 $\mu$s. Such a distribution is used to find the 
      leading edge of the plasma liner which, in this case, is 76 cm.
    } % Description of the figure and the figure will be auto-numbered
    \label{fig:ram_pres_radial} % must be after caption
    %\end{center}
 \end{figure}

    Before discussing  details of the plasma liner evolution, we introduce 
    definitions of {\it the liner leading edge and the spherical slice}. Let's take  
    a complete 3D distribution of the plasma ram pressure $\rho v^2$ at any 
    time during the simulation, where $\rho$ is the plasma density and $v$ is the 
    velocity magnitude, and average it in angular coordinates within a spherical 
    coordinate system that has its origin in the PLX chamber center. This results 
    in a radial distribution of the averaged ram pressure with respect to the 
    chamber radial coordinate. An example of such a radial distribution for 6 jets 
    at time 17 $\mu s$ is shown in Figure \ref{fig:ram_pres_radial}. We define the 
    {\it liner leading edge} $R_L$ as the radial coordinate of the maximum value 
    of the averaged ram pressure. In particular, the leading edge of the liner is 
    76 cm in Figure \ref{fig:ram_pres_radial}.

\begin{figure}[h] % flow mode; [H]: fix the position of tables, graphs
      \includegraphics[width=0.4\textwidth] 
              {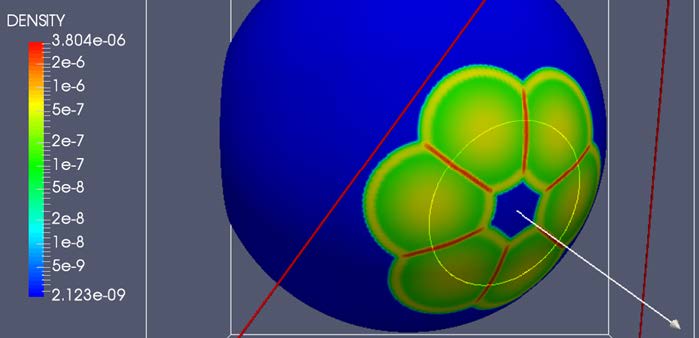} % size and path
    \caption{Density distribution on a spherical slice in the leading edge of a plasma liner, formed by 6 plasma jets.
      \raggedright
          } 
    \label{fig:sph_slice} 
\end{figure}

  We find it convenient to analyze typical states in the liner,  its non-uniformity, as well as some global properties, by using distributions of hydrodynamic states on spherical sliced drawn through the leading edge of the liner. An example of the density distribution in the leading edge of a plasma liner, demonstrating shock waves, is shown in Figure \ref{fig:sph_slice}.  In the remainder of the paper, we will show states on spherical slices in the form of 2D plots (ignoring the curvature of the spherical slice).   

\begin{figure}[h!]
    \centering
      \begin{subfigure}{0.4\textwidth} % subfigure
      \centering
        \includegraphics[width=0.7\textwidth]
          {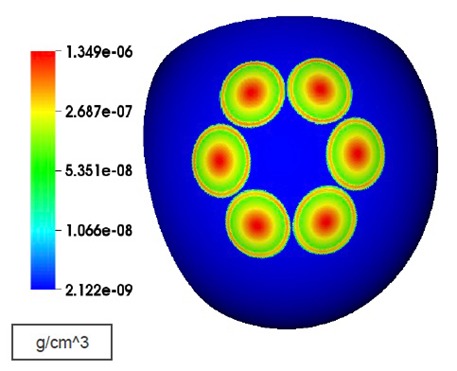}
      \caption{Before primary shocks, $t = 22 \mu s$, $R_L = 58.5$ cm.} 
      \label{fig:main_phases_a}
      \end{subfigure} 
       \begin{subfigure}{0.4\textwidth} % subfigure
      \centering
        \includegraphics[width=0.8\textwidth]
          {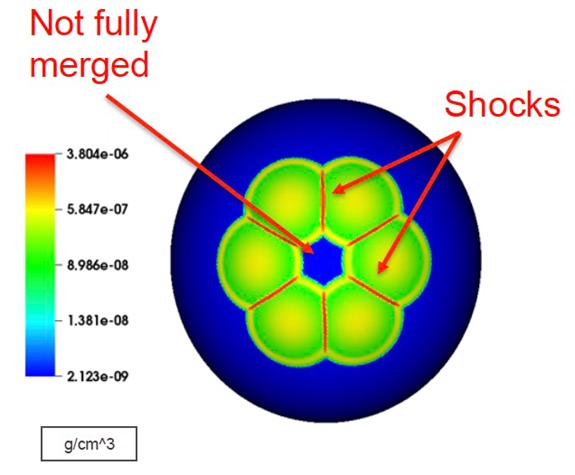}
      \caption{Primary shocks, $t = 30 \mu s$, $R_L = 30$ cm.} 
      \label{fig:main_phases_b}
      \end{subfigure} 
       \begin{subfigure}{0.4\textwidth} % subfigure
      \centering
        \includegraphics[width=0.8\textwidth]
          {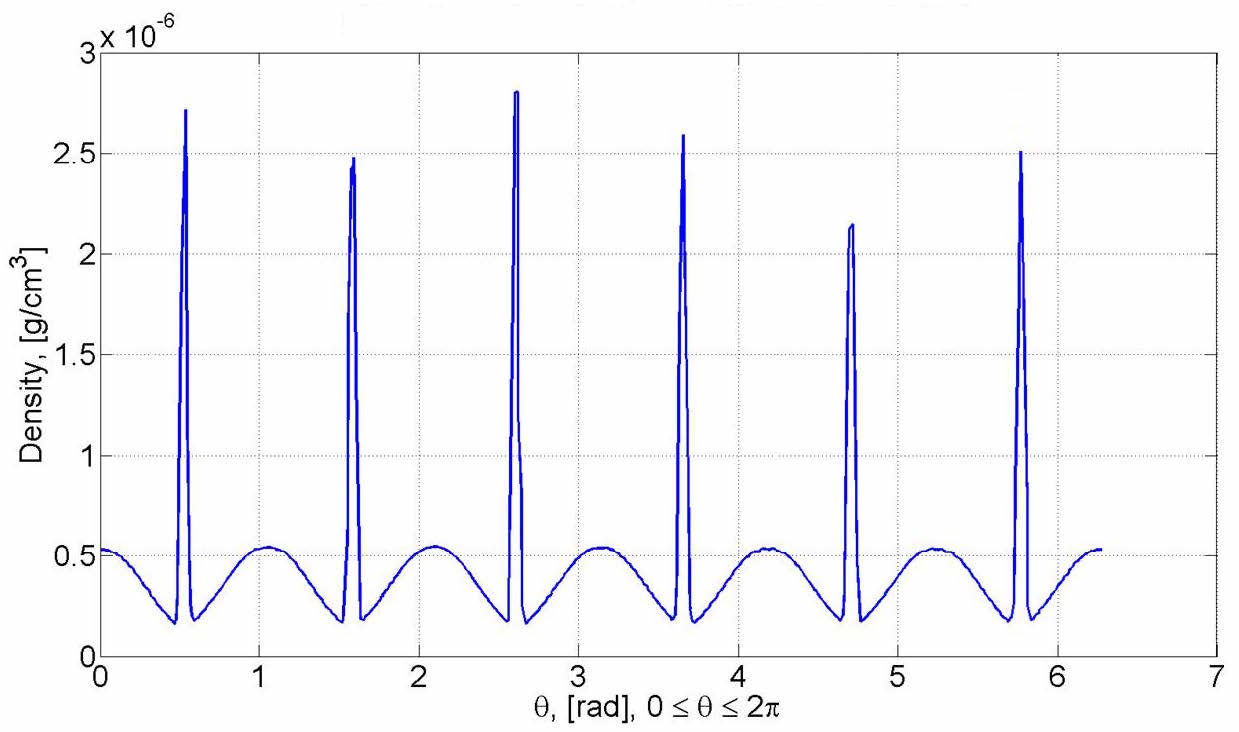}
      \caption{Density along a circular line.} 
      \label{fig:main_phases_b2}
      \end{subfigure} 
       \begin{subfigure}{0.4\textwidth} % subfigure
      \centering
        \includegraphics[width=0.7\textwidth]
          {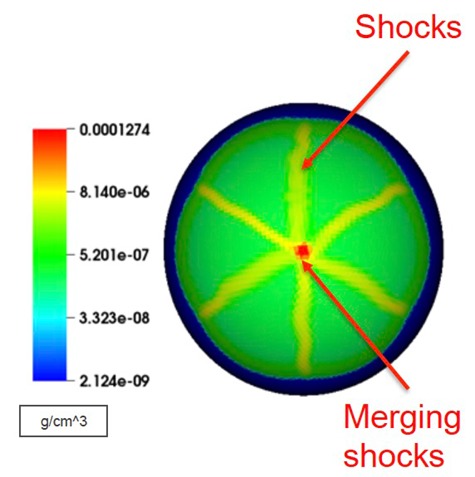}
      \caption{Secondary shocks, $t = 36 \mu s$, $R_L = 7$ cm.} 
      \label{fig:main_phases_c}
      \end{subfigure} 
    \caption{Main phases of plasma jet merger and liner formation process.
      \raggedright
          } 
    \label{fig:main_phases} 
\end{figure}

  Having defined the liner leading edge and the spherical slice, we discuss qualitatively the main phases of the plasma jet merger and liner formation process using idealized simulations with identical initial states of plasma jets.  At   time 22 $\mu$s, corresponding to $R_L = 58.5$ cm,  plasma jets are positioned closely to each other but still beyond their  interaction distance (Figure \ref{fig:main_phases_a}).
 At this stage,   the highest values of density and ram pressure are located in the 
    center of each plasma jet. When $R_L = 30$ cm at time 30 $\mu$s, plasma jets collide with each other, resulting in the formation of oblique shock waves (Figure \ref{fig:main_phases_b}).
    Locations of the highest values of density move into planes 
    between the interacting plasma jets, i.e. into the primary shock waves.
    A 1D distribution of density along a circular line shown in Figure \ref{fig:sph_slice}, drawn through shock waves and the jet centers, is depicted in
    Figure \ref{fig:main_phases_b2}). We have verified  that the density and pressure jumps are in agreement with the theory of oblique shock waves \cite{Courant}. Their large jumps are explained by low values 
    of the adiabatic index $\gamma \sim 1.2$ caused by atomic processes (ionization) and radiation in the location of shock waves. In addition, the single-fluid approximation, used in all simulations,
    does not resolve small interpenetration of edges of plasma jets and possibly over-predicts the strength of shock waves.
    
    When $R_L = 7$ cm at 
    time 36 $\mu$s, the interaction of primary shock waves is observed  (Figure \ref{fig:main_phases_c}).    Highest values of density are now located along the central axis of the group of 6 jets, corresponding to the center of the secondary shock wave. Density values in the secondary shock exceed those 
in  the primary shock waves and in the jet centers.  Such differences in states between plasma jet central regions and oblique shocks  is the main cause of non-uniformities in plasma liners. For idealized cases with identical initial jet states, cascades of oblique shock waves are  observed in simulations with any number of jets.   

\subsection{Effect of jet mass and timing variations}
     
\begin{figure}[h] % flow mode
  \begin{center}
	\includegraphics[width=0.45\textwidth, height=0.34\textheight]%[width=0.38\textwidth]
	  {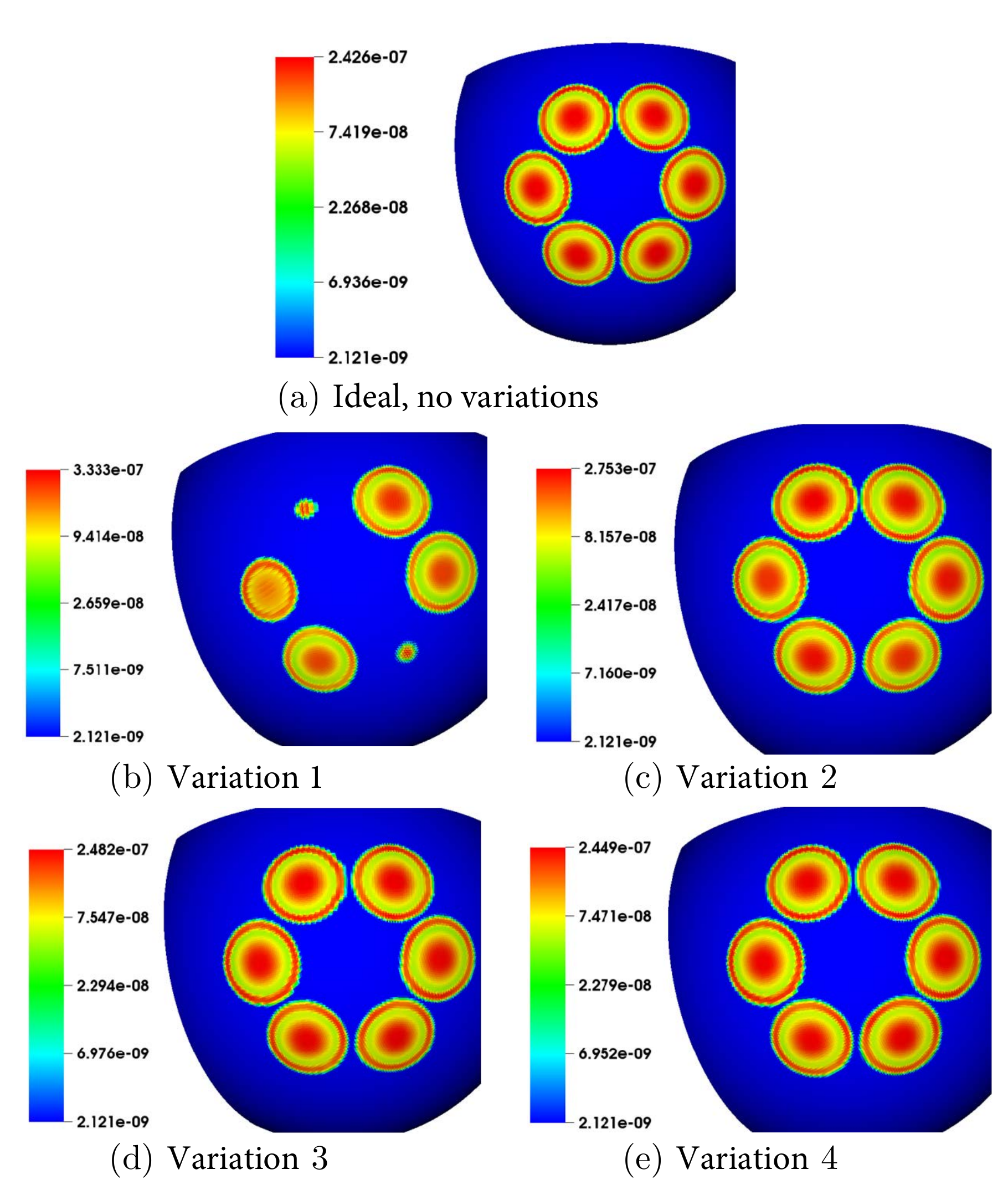} % size and path
	\caption{
	  Comparison of density profiles (g/cm$^3$) on liner leading edges with 
       with $R_L=$ 58.5 cm at 22 $\mu$s for simulations with identical initial conditions  (a) and simulations described in Table \ref{tab:perturbed_set} (b-e).
      \raggedright
	}
    \label{fig:perturb_den_22mus_2} % must be after caption
  \end{center}
  \end{figure}

\begin{figure} [h]% flow mode
  \begin{center}
	\includegraphics[width=0.45\textwidth, height=0.34\textheight]%[width=0.38\textwidth]
	  {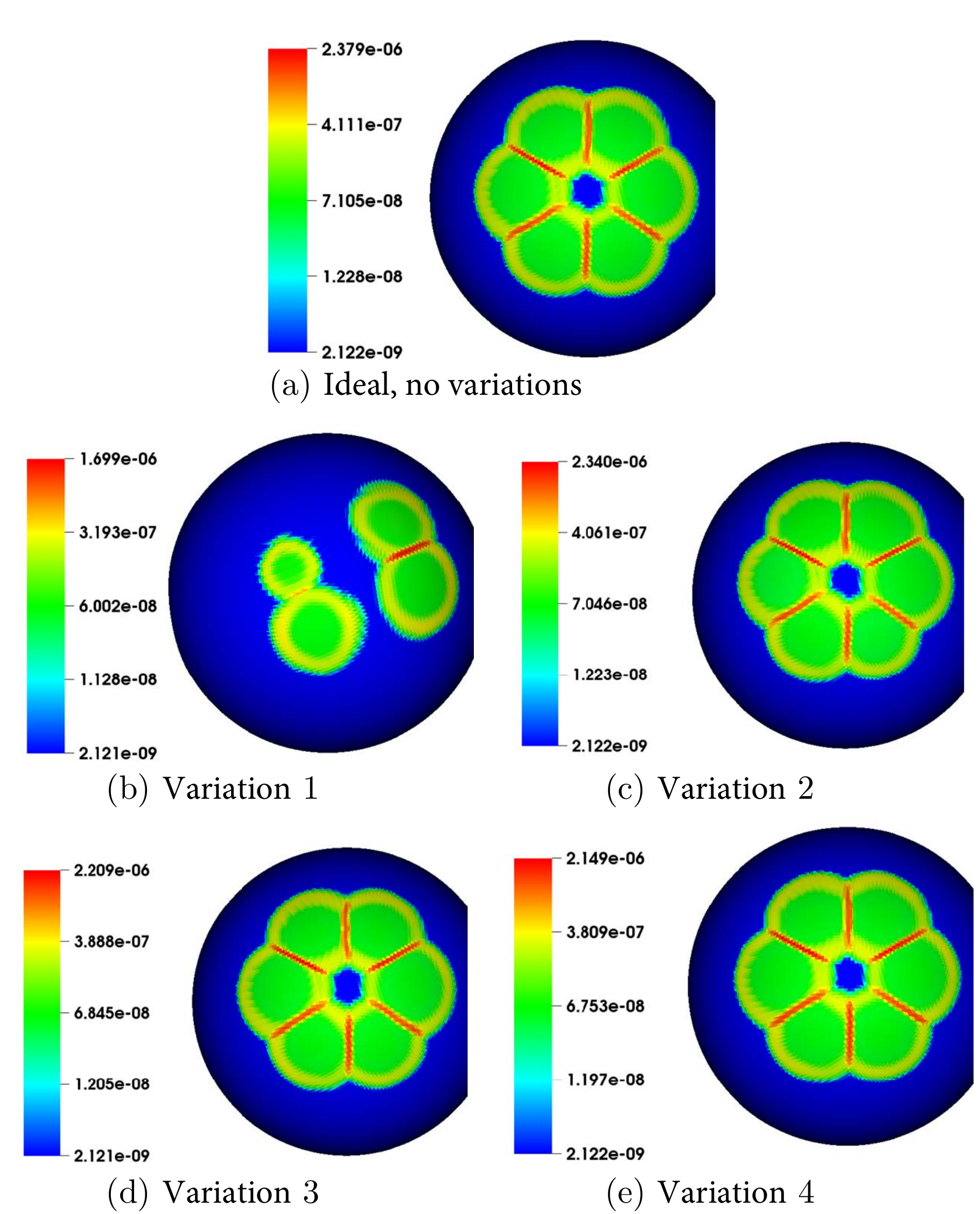} % size and path
	\caption{
 	  Comparison of density profiles (g/cm$^3$) on liner leading edges with 
       with $R_L=$ 30 cm at 30 $\mu$s for simulations with identical initial conditions  (a) and simulations described in Table \ref{tab:perturbed_set} (b-e).
      \raggedright
	}
    \label{fig:perturb_den_30mus_2} % must be after caption
  \end{center}
  \end{figure}

\begin{figure} [h] % flow mode
  \begin{center}
	\includegraphics[width=0.45\textwidth, height=0.34\textheight]%[width=0.38\textwidth]
	  {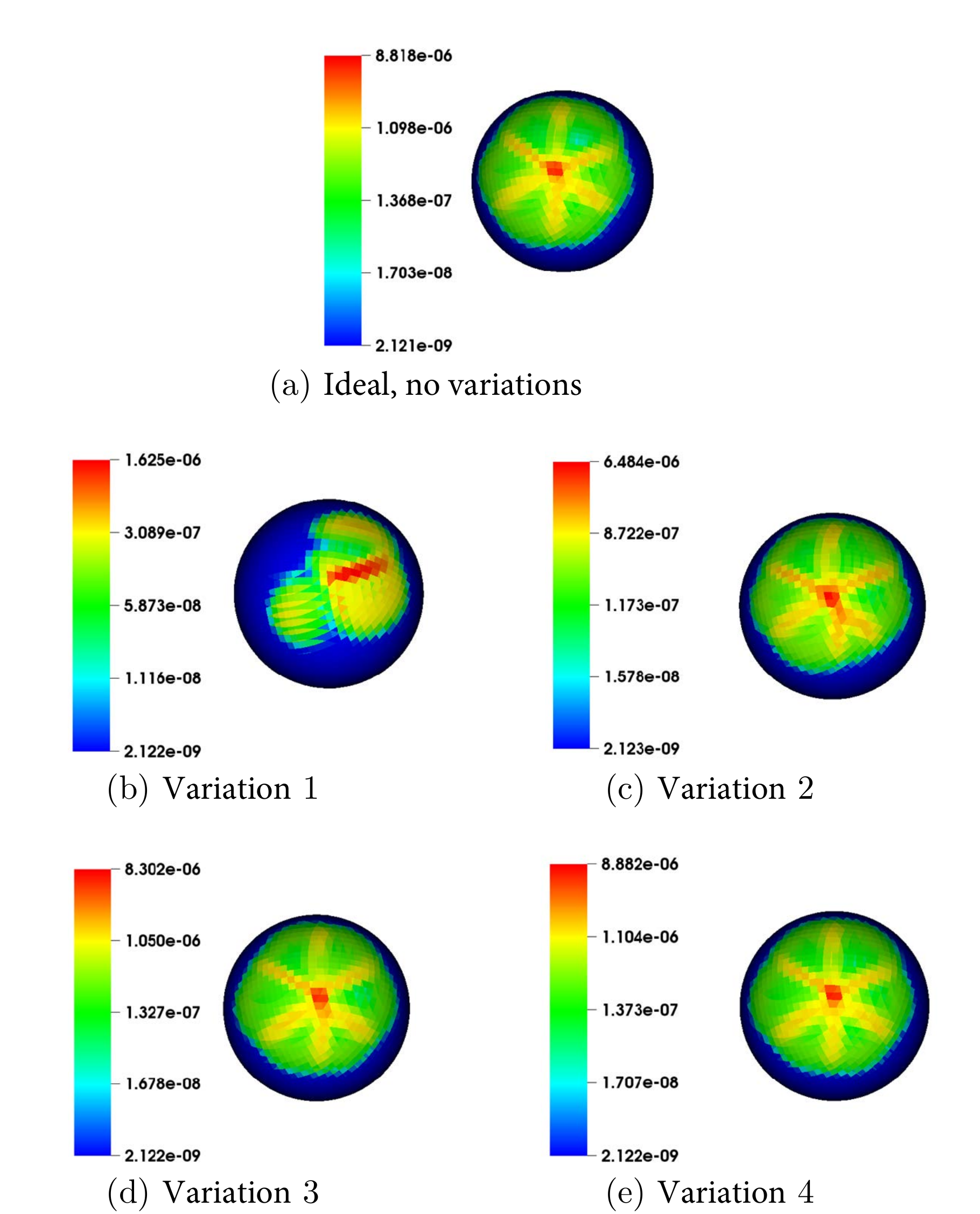} % size and path
	\caption{
 	  Comparison of density profiles (g/cm$^3$) on liner leading edges with 
       with $R_L=$ 7 cm at 36 $\mu$s for simulations with identical initial conditions  (a) and simulations described in Table \ref{tab:perturbed_set} (b-e).
      \raggedright
	}
    \label{fig:perturb_den_36mus_2} % must be after caption
  \end{center}
  \end{figure}

We now discuss the influence of experimental variations on the shock-wave structure and global properties of plasma liners. 
In the 1st experimental scenario, described in Table \ref{tab:perturbed_set}, the distribution of density states on spherical slices are 
shown in  Figures \ref{fig:perturb_den_22mus_2}(b)(c)(d)(e) - \ref{fig:perturb_den_36mus_2}(b)(c)(d)(e).
For comparison, stated from idealized simulations with identical initial conditions are shown in Figures \ref{fig:perturb_den_22mus_2}(a) - \ref{fig:perturb_den_36mus_2}(a). Due to larger changes in jet masses in the "Variation 1" simulation, combined with large time variations, 
some plasma jets did not start interacting with other jets on $R_L = 58.5$ cm at time 
    22 $\mu$s. This caused a significantly different structure of shock waves at 30 $\mu s$. 
    However, when errors in the gunshot time were reduced from  $\pm 2\mu s$ to $\pm 100$ ns and then 
    to $\pm 50$ ns, simulation data corresponding to Variations 2, 3, and 4 demonstrate that the structure of primary shock waves at 30 $\mu s$ remains almost identical to the ideal case.  We also see that results are more sensitive to shot-time variations compared to jet-mass variations.

Later stages of the liner implosion are more sensitive to small values of experimental variations of initial conditions. As expected, the structure of shock waves is completely different to the unperturbed case for Variation 1  at 36 $\mu s$, when the leading edge of the liner implodes to the radius of 7 cm. But there are also visible changes of shock waves for simulations corresponding to Variations 2 - 4.  
This simulation prediction was confirmed by experimental observations described in the next section.
Distributions of other states on spherical slices such as ram pressure, hydrostatic pressure, and temperature show similar features. 

      %%%%%%%%%%%%%%%%
  \begin{figure}[h!] % flow mode
  \begin{center}
	\includegraphics[width=0.45\textwidth, height=0.34\textheight]%[width=0.48\textwidth]
	  {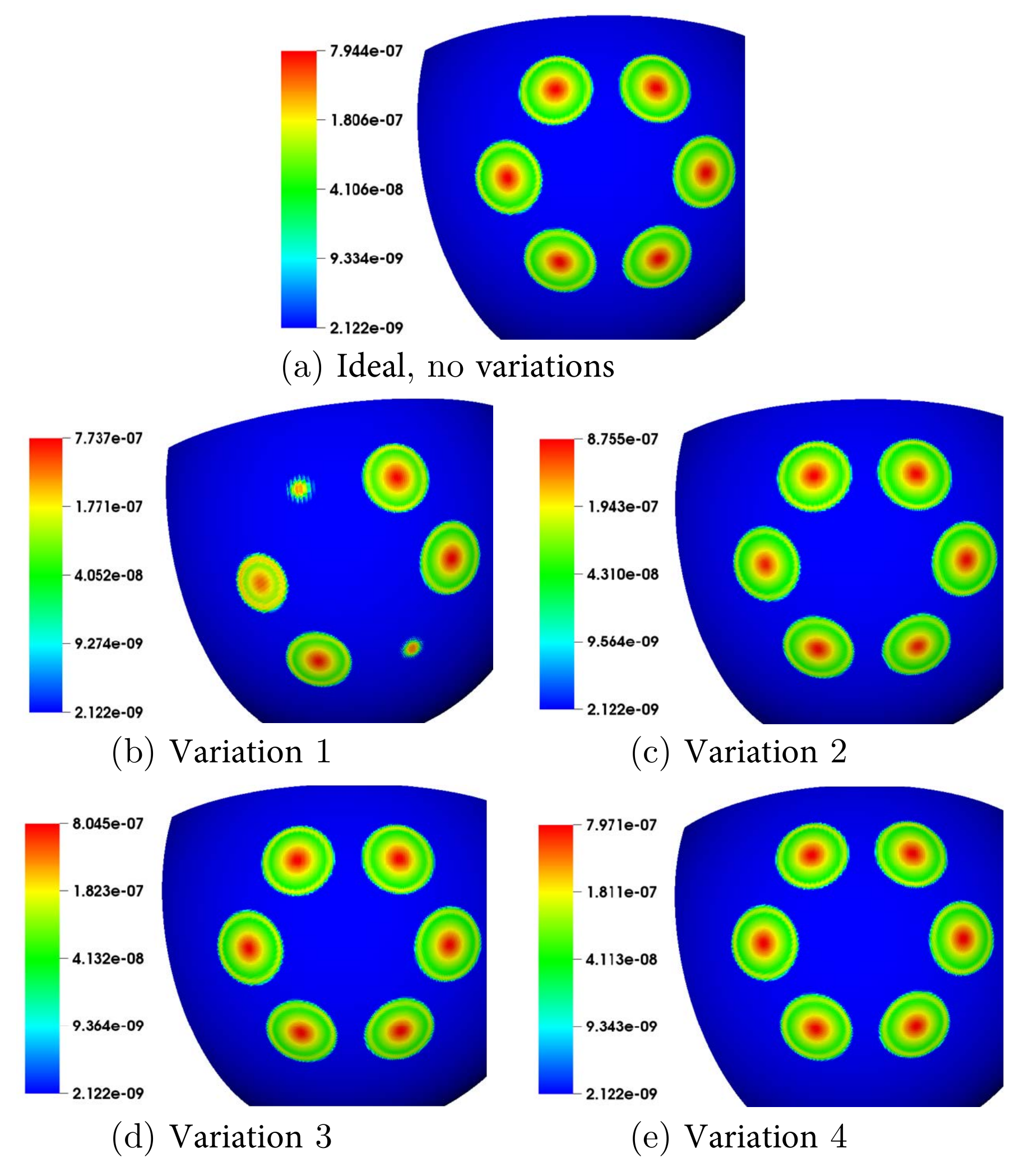} % size and path
	\caption{
	  Comparison of density profiles (g/cm$^3$) on liner leading edges with 
       with $R_L=$ 76 cm at 17 $\mu$s for simulations with identical initial conditions (a) and simulations described in Table \ref{tab:perturbed_set} (b-e).
      \raggedright
	}
    \label{fig:perturb_den_17mus_all} % must be after caption
  \end{center}
  \end{figure}

  % end figure %%%%%%%%%%%%%%%%%%%%%%%%
  %%%%%%%%%%%%%%%%%%%%%%%%%%%%%%%%%%%%% 
         
  %%%%%%%%%%%%%%%%%%%%%%%%%%%%%%%%%%%%%
  % begin figure: flow %%%%%%%%%%%%%%%%
  % end figure %%%%%%%%%%%%%%%%%%%%%%%%
  %%%%%%%%%%%%%%%%%%%%%%%%%%%%%%%%%%%%% 
  
  We now investigate how experimental variations affect global properties of plasma liners. 
 The averaged Mach number and ram pressure of a liner are the most important global properties that characterize the ability of the liner to compress plasma targets.  In particular, it is very important to maintain high Mach numbers of liners during their evolution in order to allow for high values of fusion energy gain \cite{LangHsu17}.  Of course, this can be achieved only for a spherically symmetric array of plasma jets. In the case of the self-implosion of  six jets, jets continue their motion beyond the chamber center. Nevertheless, the dependence of this global characteristic on experimental variations is important for a better understanding of the liner performance.

\begin{figure}[h] % flow mode
  \begin{center}
	\includegraphics[width=0.45\textwidth]
	  {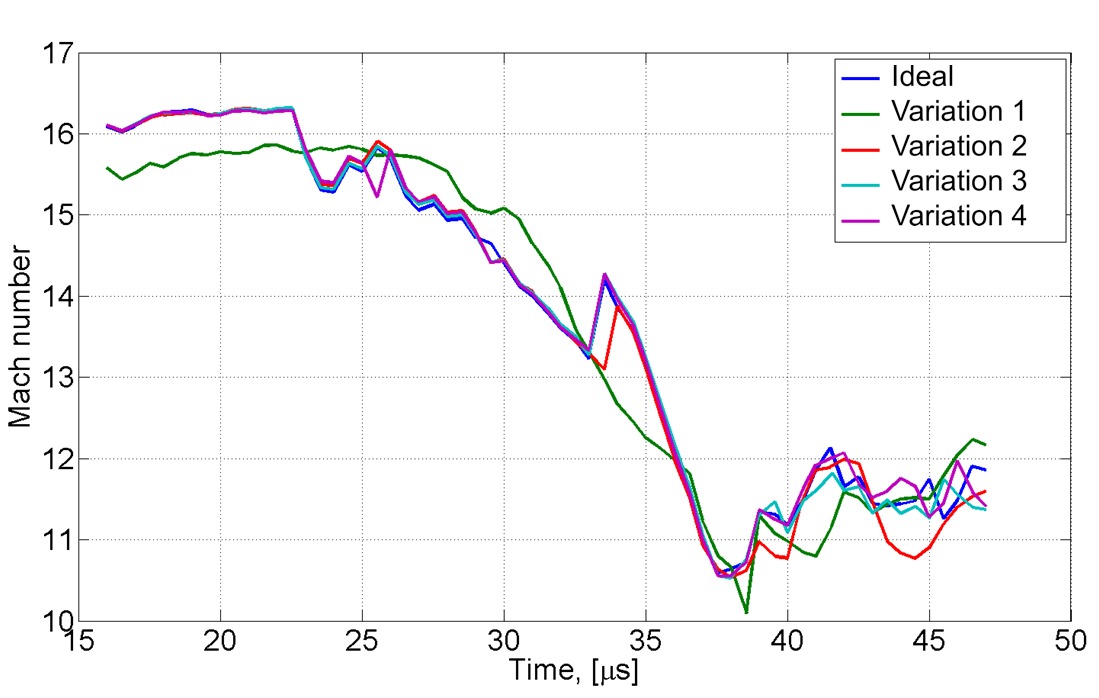} % size and path
	\caption{
	  Evolution of averaged Mach number in imploding liners in simulations with identical initial conditions  and with experimental variations described in Table \ref{tab:perturbed_set}.
      \raggedright
	}
    \label{fig:mach_1} % must be after caption
  \end{center}
  \end{figure}

\begin{figure}[h] % flow mode
  \begin{center}
	\includegraphics[width=0.45\textwidth]
	  {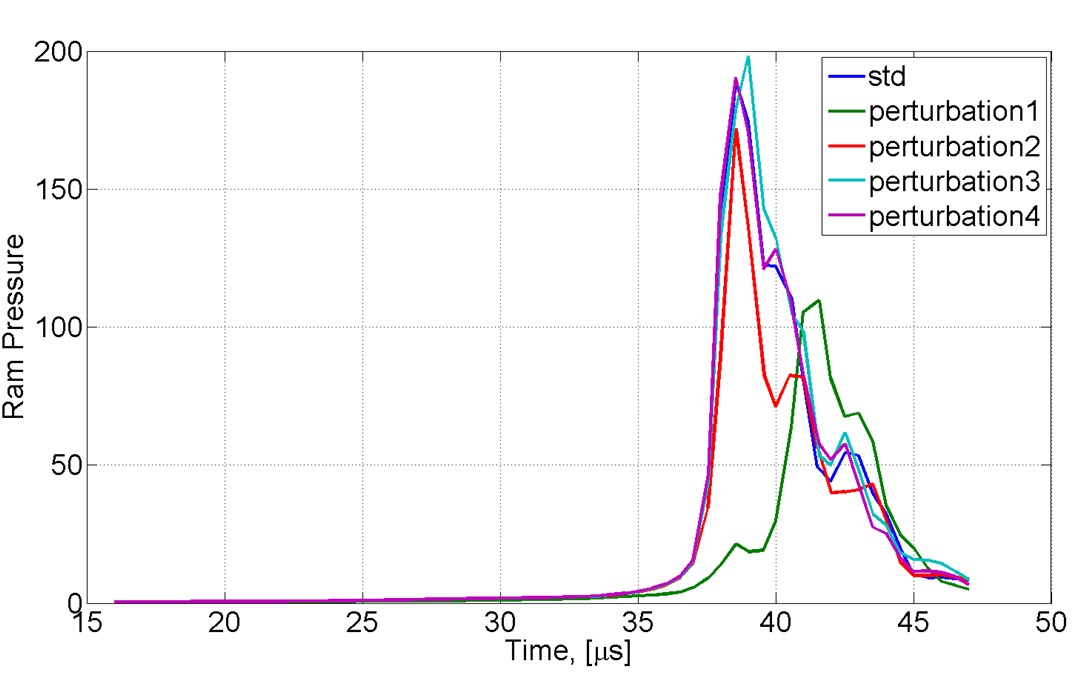} % size and path
	\caption{
Evolution of averaged ram pressure on the leading edges of imploding liners in simulations with identical initial conditions  and with experimental variations described in Table \ref{tab:perturbed_set}.	
	        \raggedright
}
    \label{fig:ram_press_1} % must be after caption
  \end{center}
  \end{figure}

In Figures \ref{fig:mach_1} and  \ref{fig:ram_press_1} we plotted the evolution of the averaged Mach number and the averaged ram pressure in imploding liners in simulations with identical initial conditions and with experimental variations described in Table \ref{tab:perturbed_set}. We observe that these variations have a very small effect on the averaged Mach number. For smaller values of variations (Variation 2 - Variation 4), all plots are very close to each other, at least before the leading edges reach the chamber center at 38 $\mu s$, and exhibit a typical increase of the Mach number caused by the primary shock waves at 33 - 34 $\mu s$. For the largest value of the experimental variations (Variation 1), the overall behavior is still similar, but the increase of Mach number due to the formation of primary shocks is not observed: the shock structure is very different in this case as different pairs of shocks form at different times. Similar conclusions can also be obtained about the evolution of the averaged ram pressure on the leading edge of the liner: three lines corresponding to smaller variations are close to each other and to the line representing the simulation with identical initial conditions. The simulation with the largest variations (Variation 1) reaches a much smaller maximum value of the averaged ram pressure  because jets  reach the chamber center at significantly different times.
  
\begin{figure}[h!] % flow mode
  \begin{center}
	\includegraphics[width=0.5\textwidth]
	  {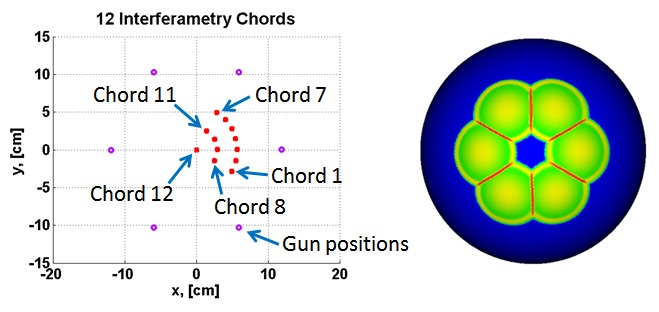} % size and path
	\caption{
	  6 jets and 12 end-on interferometry chords.
      \raggedright
	}
    \label{fig:interf_initial} % must be after caption
  \end{center}
  \end{figure}

 A multi-chord interferometer is installed in the PLX chamber to probe the liner plasma. The interferometer measures the phase shift that is proportional to the integral of the electron number density along the interferometer chords. Figure \ref{fig:interf_initial} gives a schematic of the interferometer chord placements. For a reference, a density distribution on a liner spherical slice is shown on the right at the correct angle to the interferometer schematic.  This implies, for example, that chords 1 and 5 should measure signals from primary shock waves, chords 3 and 7 should probe plasma along jet centers, and the chords closer to the center should detect the secondary shock waves or combinations of primary and secondary shocks. At present, chords 7  - 12 are not installed in the PLX  chamber, but we present simulation data along them as well.
 
\begin{figure}[h] % flow mode
  \begin{center}
	\includegraphics[width=0.5\textwidth]
	  {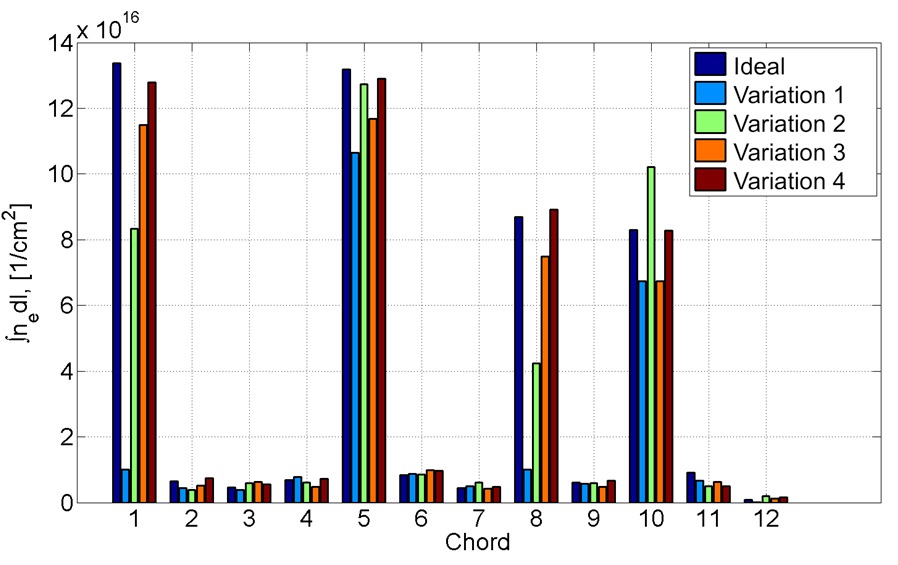} % size and path
	\caption{
	  Multi-chord interferometer data through chord 1 to chord 12 
      described at 30 $\mu$s.
      \raggedright
	}
    \label{fig:perturb_interf_30mus} % must be after caption
  \end{center}
  \end{figure}

Figure \ref{fig:perturb_interf_30mus} presents an example of computed signals along the interferometer chords using simulation data from initial states as described in Table \ref{tab:perturbed_set}. The purpose of this example is not to give a comprehensive information on all interferometer readings at all time but rather demonstrate their sensitivity to the variations of initial jet parameters. Our results show that interferometer signals are quite sensitive to 
experimental variations, especially at earlier times, due to the fact that shock waves form initially thin structures in plasma liners: if the location of a primary shock  is shifted, the corresponding interferometer signal changes significantly (see interferometer signals for Variation 1 case). But if the structure of shock waves is more stable, interferometer signals on chords 1 and 3  may change by 10 \% - 30\%. Signals on the chords corresponding to jet centers depend on jet lengths, and thus are stronger for longer jets used in experiments.

  % end figure %%%%%%%%%%%%%%%%%%%%%%%%
  %%%%%%%%%%%%%%%%%%%%%%%%%%%%%%%%%%%%%
  
\subsection{Effect of jet mass/velocity variations}
   
\begin{figure}[h] % flow mode
  \begin{center}
	\includegraphics[width=0.5\textwidth]
	  {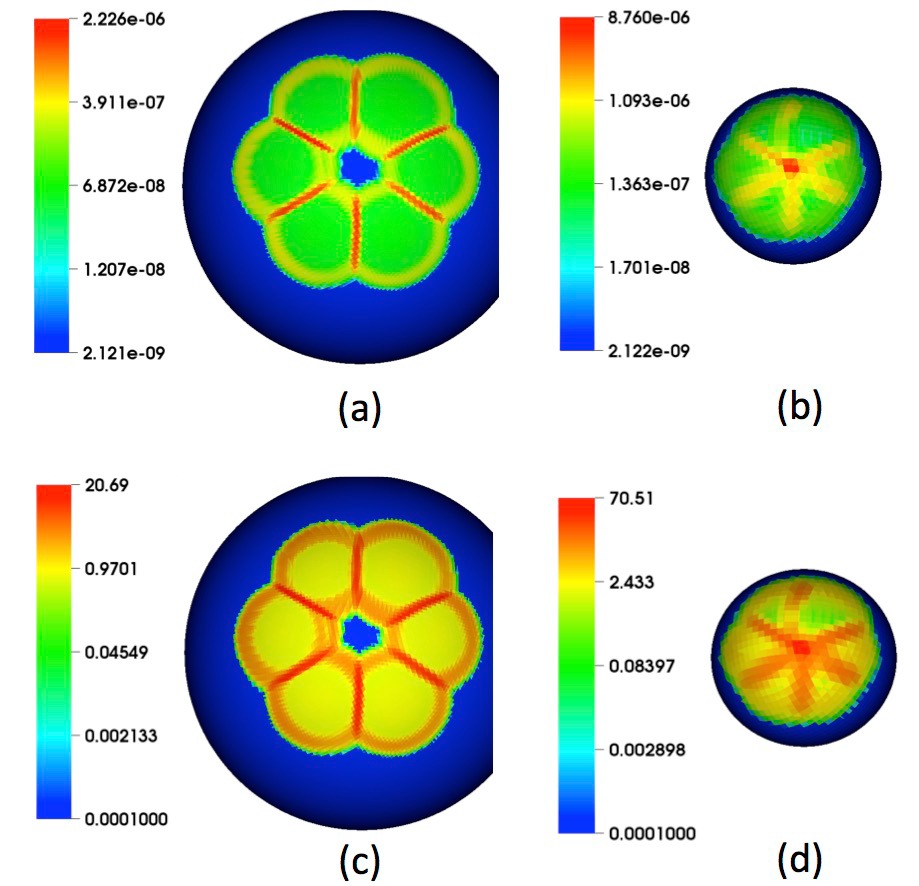} % size and path
	\caption{Density (g/cc, top row of images) and ram pressure (bar, bottom row) on spherical slices of the plasma liner at 30 $\mu s$ (left column) and   36 $\mu s$ (right column). Simulation was initialized as described in Table \ref{tab:perturbed_second_set}.
      \raggedright
	}
    \label{fig:pert_2} % must be after caption
  \end{center}
  \end{figure}

In this section, we briefly summarize simulation results for the second experimental scenario, in which the initial mass of argon in plasma guns changes by $\pm 5 \%$, but the kinetic energy of jets remains the same and all jets are shot perfectly synchronously (see Table \ref{tab:perturbed_second_set}).  Figure \ref{fig:pert_2} depicts density (images (a) and (c)) and ram pressure (images (c) and (d)) on spherical slices of the plasma liner at 30 $\mu s$ (left column) and   36 $\mu s$ (right column). As before, we observe that the structure of primary shock waves is unchanged under small variations, but the later stages of the evolution are more sensitive to variations (see images at 36 $\mu s$). 

\begin{figure}[h] % flow mode
  \begin{center}
	\includegraphics[width=0.5\textwidth]
	  {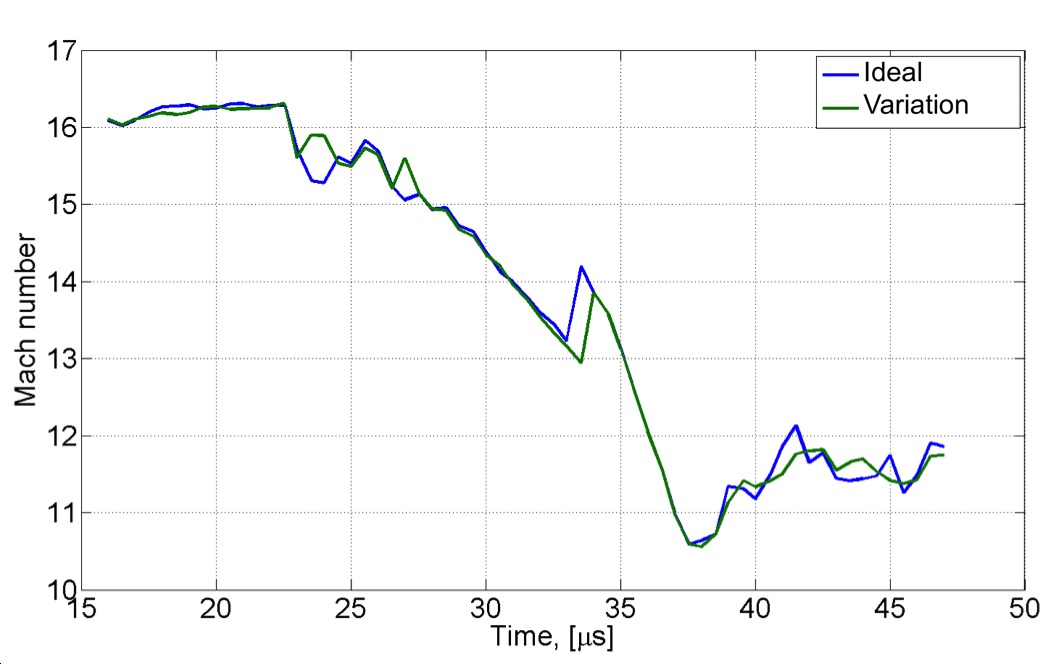} % size and path
	\caption{Averaged Mach number evolution in imploding liners with identical initial conditions (blue line) and with variations described in Table \ref{tab:perturbed_second_set} (green line).
      \raggedright
	}
    \label{fig:mach_number_2} % must be after caption
  \end{center}
  \end{figure}

\begin{figure}[h] % flow mode
  \begin{center}
	\includegraphics[width=0.5\textwidth]
	  {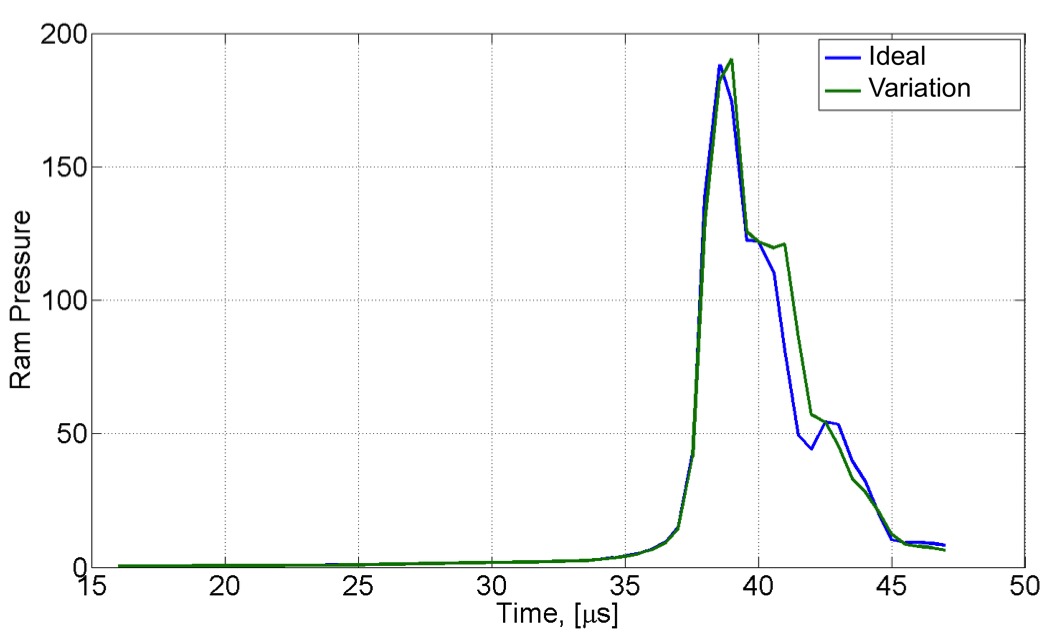} % size and path
	\caption{Evolution of the averaged ram pressure in the liner leading edge with identical initial conditions (blue line) and with variations described in Table \ref{tab:perturbed_second_set} (green line).
	      \raggedright
}
    \label{fig:ram_pres_2} % must be after caption
  \end{center}
  \end{figure}

\begin{figure}[h]
    \centering
       \begin{subfigure}{0.5\textwidth} % subfigure
      \centering
        \includegraphics[width=0.9\textwidth]
          {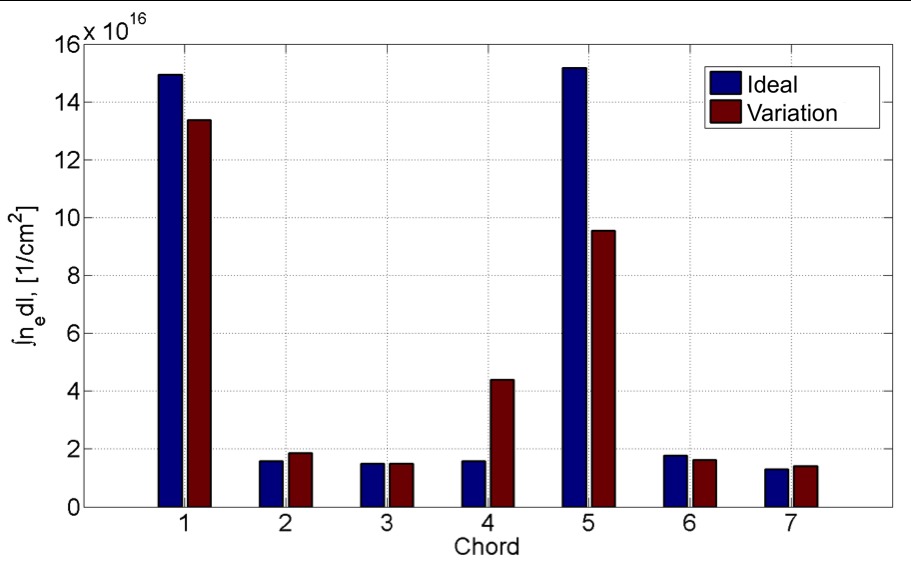}
      \caption{30 $\mu s$} 
      \label{fig:interf_2_1}
      \end{subfigure} 
       \begin{subfigure}{0.5\textwidth} % subfigure
      \centering
        \includegraphics[width=0.9\textwidth]
          {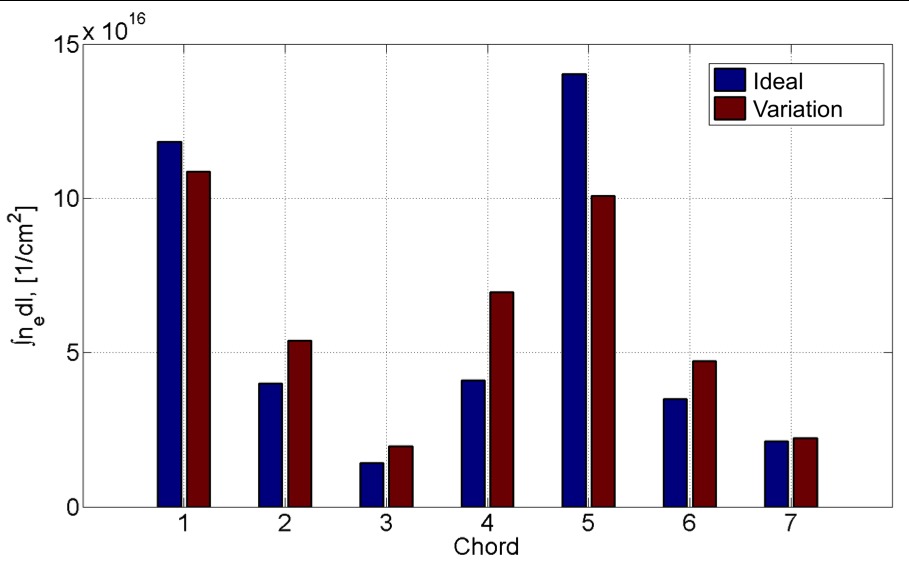}
      \caption{36 $\mu s$} 
      \label{fig:interf_2_2}
      \end{subfigure} 
	\caption{Interferometry signals (line integrals of the electron density along interferometer chords) at (a) 30 $\mu s$ and  (b) 36 $\mu s$ in imploding liners with identical initial conditions (blue color) and with variations described in Table \ref{tab:perturbed_second_set} (red color).
      \raggedright
}
    \label{fig:interf_2} % must be after caption
  \end{figure}

 Figure \ref{fig:mach_number_2} shows the averaged Mach number evolution in imploding liners with identical initial conditions (blue line) and with variations described in Table \ref{tab:perturbed_second_set} (green line). This result as well as the evolution of the averaged ram pressure (Figure \ref{fig:ram_pres_2}) in the liner leading edge are also consistent with previous simulations and show that the global properties of plasma liners are not sensitive to small experimental variations. Despite the relative stability of primary shock waves, interferometry signals corresponding to primary shocks  (chords 1 and 5 in Figure \ref{fig:interf_2}) are sensitive to small experimental variations at early time because of small displacements of narrow shock wave structures. The discrepancy between interferometry signals reduces with time as oblique shock regions widen (see Figure \ref{fig:interf_2_2}).
 
\section{Comparison of Experimental Results with Simulations}

\begin{figure}[h]
    \centering
       \begin{subfigure}{0.9\textwidth} % subfigure
      \centering
        \includegraphics[width=0.8\textwidth]
          {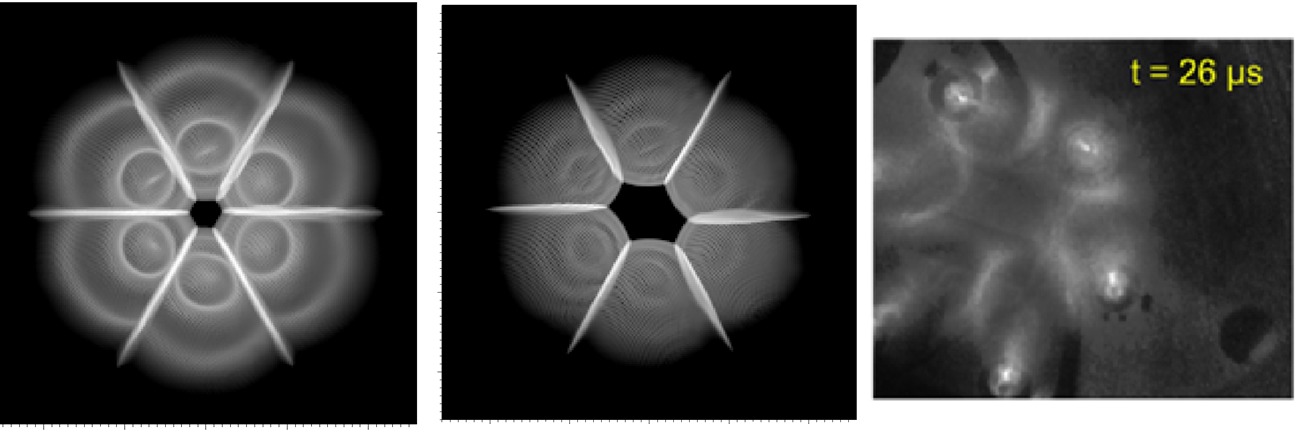}
      \caption{26 $\mu s$} 
      \label{fig:c3}
      \end{subfigure} 
       \begin{subfigure}{0.9\textwidth} % subfigure
      \centering
        \includegraphics[width=0.8\textwidth]
          {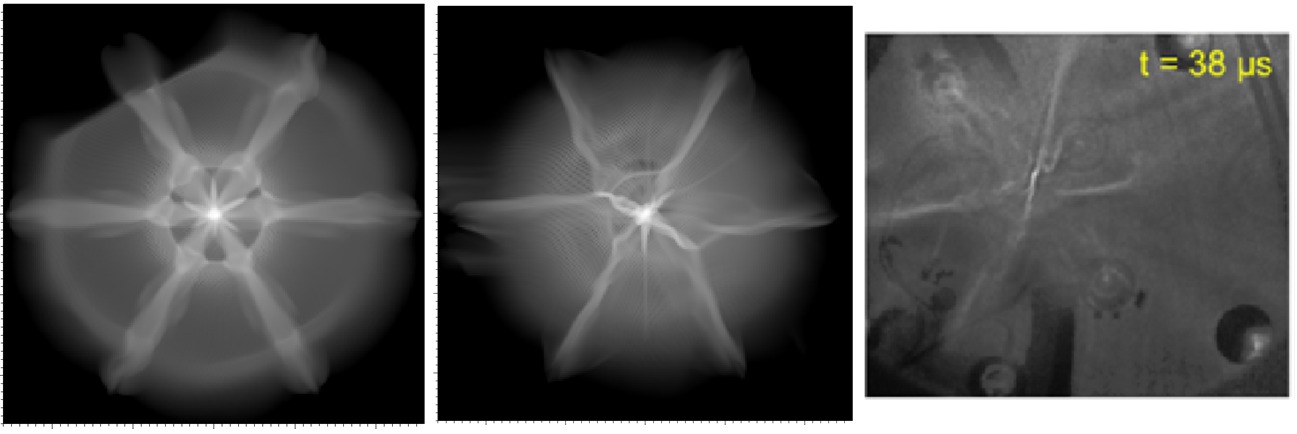}
      \caption{38 $\mu s$} 
      \label{fig:c3}
      \end{subfigure} 
       \begin{subfigure}{0.9\textwidth} % subfigure
      \centering
        \includegraphics[width=0.8\textwidth]
          {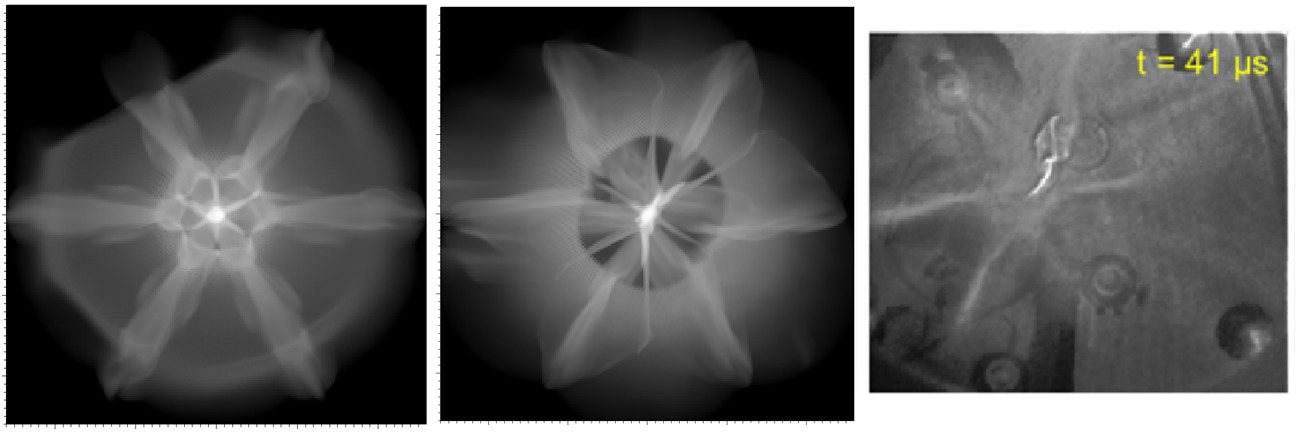}
      \caption{41 $\mu s$} 
      \label{fig:c6}
      \end{subfigure} 
    \caption{Comparison of simulations with experimental CCD camera images.  The left image for every time represents the distribution of the integrated radiation power from the simulation initialized as in Table \ref{tab:perturbed_second_set} and the middle image represents the same quantity from the simulation initialized as in Variation 1 entry of Table \ref{tab:perturbed_set} (more extreme case of variation of initial parameter). Experimental results are shown in the right images.
      \raggedright
          } 
    \label{fig:exp_comp_1} 
\end{figure}

In this Section we confirm our simulation predictions by comparing numerical results with experimental data. High-quality CCD images of
imploding liners formed by six plasma jets have been obtained in recent experiments. Comparison of CCD images with simulations is in itself a difficult task as experiments and simulations operate with very different sets of quantities. Using simulation data, we approximate the experimental CCD images by assuming that each pixel is proportional to the line integral of the radiation power across the entire simulation domain in the normal direction to the CCD camera element. This approach ignores the spectral sensitivity of the CCD camera element as well as absorption along the light path. With such an interpretation of the CCD images, comparison of simulations with experiments is presented in Figure \ref{fig:exp_comp_1}.
The left image for every time represents the integrated radiation power  distribution obtained from the simulation initialized as in Table \ref{tab:perturbed_second_set} and the middle image represents the same quantity obtained from the simulation initialized as in Variation 1 entry of Table \ref{tab:perturbed_set}. Experimental results are shown in the right images.
 
 \onecolumngrid
 
 %%%%%Comparison with experiments: density
\twocolumngrid

Experimental data confirm main features of the detailed liner structure predicted by simulations. At 26 $\mu s$, we observe the formation of primary shock waves. The brighter spots in the centers of six jets in experimental images can be explained by longer jets used in experiments compared to 10 cm long simulated jets. The primary shocks are observed in both simulations images at 26 $\mu s$, but the right image (larger variations) is a closer match to the experimental image.
As the liner continues to implode, the structure of primary shock waves remain stable. At later times (38 and 41 $\mu s$), secondary shock waves are clearly visible in the center of both experimental and simulation images. While  the shape of secondary shocks  in the simulation with smaller variations of initial conditions remains symmetric star-like, only strongest shocks are clearly visible in simulations with large initial variations and in experiments. In addition, most dominant shocks may change the direction in time (secondary shocks created by other pairs of primary shocks may start dominating at later time), as we see in simulations  with large initial variations (top right images) at  38 and 41 $\mu s$. Such a dynamics of secondary shocks depends on a specific realization of random initial settings and does not always occur in simulations.

Experimental images provide confirmation of our prediction that the primary shock wave structure is much less sensitive to the experimental variations compared to the structure of secondary shocks.  As maximum values of the variation of jet parameters in real experiments are unknown, comparison to simulations may provide useful information on their amplitude.

Experimental data on interferometry is also available  \cite{PLX18}. New and improved interferometer data with significantly better jet-to-jet mass balance than reported in \cite{PLX18} will be reported in a separate paper. Both simulation and experimental interferometry results overlap due to their large error bars.  Large error bars of the experimental data from the multi-chord interferometer can be explained by our simulation observation that interferometer signals are sensitive to experimental errors. But because CCD images and interferometry are available only for different experimental runs, their direct comparison at this point is not justified. In addition, longer jets used in experiments lessen the amplitude of interferometry signals coming from the regions of jet bodies and shock waves.

\section{Summary and Conclusions}

We have performed simulations studies of the merger of 6 plasma jets and the formation and implosion of plasma liners at conditions that model the PLX-$\alpha$ 
experiment at Los Alamos. Sensitivity studies of the detailed structure of plasma liners and their global properties to experimental variations in plasma guns that change initial conditions of plasma jets were the primary goal of this paper. Experimentally observable synthetic
quantities have been computed using simulations data and compared with the available experimental data. Simulations predict that the primary oblique shock wave structure is preserved at small experimental variations. If initial conditions are too different in all jets (Variation 1 entry in Table \ref{tab:perturbed_set}), the topology of primary shocks is also changed. At later phases of the liner implosion, primary shocks and, especially, secondary shocks are more sensitive to experimental variations. 
The liner structure is more sensitive to the variations of jet shot times compared to the variations of jet masses.
Small displacements of shock wave structures, which are confined to relatively narrow layers of high density and pressure plasma, cause significant changes in the interferometer readings, computed using the simulation data, especially at early time. As shock wave regions widen at later time, the interferometry readings become less sensitive for small values of variations. Our studies also showed that the global properties of the plasma liners (averaged Mach number and averaged ram pressure along leading edges of plasma liners) are not very sensitive to experimental variations. 

Simulations have achieved a good agreement with experimental CCD camera images. Both CCD images and large error bars of  the experimental interferometry data confirm our simulation predictions on the detailed structure of plasma liners and its sensitivity to experimental variations. Experimental data on global properties of plasma liners will be available in the future. 

It must be emphasized that the present experiments are not intended to demonstrate fusion-worthy liner quality and uniformity, which will most likely require smaller merging angles between plasma jets and many more jets (e.g., hundreds of jets over a sphere of 3 to 4-m radius).  The purpose of this work is to benchmark our models and codes such that they may be used to investigate and help design follow-on experiments with more jets at smaller merging angles.
Future simulations will focus on properties of plasma liners formed by $4\pi$-symmetric array of plasma jets, varying from 36 to 200 jets and the compression of gaseous and plasma targets. A recently developed Lagrangian Particle method \cite{SamWang18}, a consistent and convergent 
particle-based numerical method for hydrodynamic problems and PDE's in general, will also be used in future liner simulations due to its property of continuously adaptive resolution.

{\bf Acknowledgements} This research is supported by the Accelerating Low-cost Plasma Heating and Assembly (ALPHA) 
Program of the U.S. Department of Energy's Advanced Research Projects Agency - Energy (ARPA-E).

%\section*{Bibliography}


\begin{thebibliography}{100}
%1 
 \bibitem{Thio99}
Y.C. F. Thio, E. Panarella, C.E. Knupp R.C. Kirkpatrick, F. Wysocki, P. Parks, G. Schmidt,
\newblock Magnetized target fusion in a spheroidal geometry with standoff drivers.
\newblock {\em in Current Trends in International Fusion Research II, edited by 
E. Panarella, National Research Council Canada, Ottawa, Canada}, 1999.
%2
\bibitem{Hsu12}
S.C. Hsu, T.J. Awe, S. Brockington, A. Case, J.T. Cassibry, G. Kagan,
S. J. Messer, M. Stanic, X. Tang, D. R. Welch, and F. D. Witherspoon,
Spherically imploding plasma liners as a standoff driver for magnetoinertial
fusion, IEEE Trans. Plasma Sci., vol. 40, p. 1287, 2012.
%3
\bibitem{Knupp14}
C. E. Knapp and R. C. Kirkpatrick, Possible energy gain for a plasma liner-
driven magneto-inertial fusion concept, Phys. Plasmas, vol. 21, p.
070701, 2014.
 %4
\bibitem{PLX18}
S. C. Hsu, S. J. Langendorf, K. C. Yates, J. P. Dunn, S. Brockington, A. Case, E. Cruz, F. D. Witherspoon, M. A. Gilmore, J. T. Cassibry, 
R. Samulyak, P. Stoltz, K. Schillo, W. Shih, K. Beckwith, Y. C. F. Thio,
Experiment to Form and Characterize a Section of a Spherically Imploding Plasma Liner, IEEE Trans. Plasma Sci., 46 (2018), Issue 6, 1951 - 1961. 
%5
\bibitem{Thio02} Y. C. Francis Thio and
  Ronald. C. Kirkpatrick. Magnetized target fusion driven by plasma
  liners. In Annual Meeting of the American Nuclear Society,
  Hollywood, FL, 2002.
%6
\bibitem{Parks08} P. Parks, On the efficacy of imploding plasma liners
  for magnetized fusion target compression, Phys. Plasmas, 2008,
  Vol. 15, p. 062506.
%7
\bibitem{SamParWu10}
R.~Samulyak, P.~Parks, and L.~Wu.
\newblock Spherically symmetric simulation of plasma liner driven 
magnetoinertial fusion.
\newblock {\em Phys. Plasmas}, 17:092702, 2010.

%8
 \bibitem{KimSam12}
H. Kim, R. Samulyak, L. Zhang, P. Parks, Influence of atomic processes on the implosion of plasma liners, Physics of Plasmas, 9:0827111,
2012.
%9
\bibitem{Awe11}
T.~J. Awe, C.~S. Adams, J.~S. Davis, D.~S. Hanna, and S.~C. Hsu.
\newblock One-dimensional radiation-hydrodynamic scaling studies of imploding 
spherical plasma liners.
\newblock {\em Phys. Plasmas}, 18:072705, 2011.
%10
\bibitem{LangHsu17}
S. J. Langendorf, S. C. Hsu, Semi-analytic model of plasma-jet-driven magneto-inertial fusion, Phys. Plasmas 24, 032704 (2017).
%11
\bibitem{HsuLang18}
S. Hsu, S. Langendorf,
Magnetized plasma target for plasma-jet-driven magneto-inertial fusion,
arXiv:1803.03323v2.
%12
 \bibitem{FronTier}
B. Fix, J, Glimm, X. Li, Y. Li, X. Liu, R. Samulyak, Z. Xu,
A TSTT integrated FronTier code and its applications in computational fluid physics,
Journal of Physics: Conf. Series, 16 (2005), 471 - 475.
%13
\bibitem{KimZhaSam13}
H. Kim, L. Zhang, R. Samulyak, P.Parks, 
On the structure of plasma liners for plasma jet induced  magnetoinertial fusion,
Phys. Plasmas 20, 022704 (2013).
 
%14
\bibitem{PROPACEOS}
J.J. MacFarlane, I.E. Golovkin, P.R. Woodruff,
HELIOS-CR -- A 1-D radiation-magnetohydrodynamics code with inline atomic kinetics modeling,
\newblock {\em Journal of Quantitative Spectroscopy \& Radiative Transfer}, 99 (2006) 381?397.

%15
\bibitem{Shih18}
W. Shih, 
Modeling and Simulation Studies of Plasma-Liner Formation and Implosion for the PLX-$\alpha$ Project,
Ph.D. thesis, Stony Brook University, 2018.
 
 %16
\bibitem{Courant}
R.~Courant, K.O.~Friedrichs.
\newblock {\em Supersonic Flow and Shock Waves}, 
\newblock Springer, 1991.

%17  
\bibitem{SamWang18}
 R. Samulyak, X. Wang, H.-S. Chen, 
 Lagrangian particle method for compressible fluid dynamics, 
 {\em J. Comput. Phys.}, 362 (2018),  1-19.

\end{thebibliography}
\end{document}